\documentclass[reprint,amsmath,amssymb,aps,pra,superscriptaddress]{revtex4-2}
\usepackage{makecell}
\usepackage{graphicx}
\usepackage{amsmath}
\usepackage{bm}
\usepackage{physics}
\usepackage{hyperref}
\usepackage{booktabs}
\usepackage{subcaption}
\begin{document}

\title{Noise-Adaptive Predictive Dynamical Decoupling}

\author{Ali Abu-Nada}
\affiliation{Sharjah Maritime Academy, Sharjah, UAE}
\author{Subhashish Banerjee}
\affiliation{Indian Institute of Technology Jodhpur-342030, India}
\author{Lian-Ao Wu}
%\email{}
\affiliation{Department of Physics, University of the Basque Country UPV/EHU, 48080 Bilbao, Spain}
\affiliation{IKERBASQUE, Basque Foundation for Science, Bilbao 48011, Spain}
\affiliation{EHU Quantum Center, University of the Basque Country UPV/EHU, Leioa, Biscay 48940, Spain}

\date{\today}
\begin{abstract}

Protecting quantum coherence against realistic environmental noise remains one of the fundamental obstacles to scalable quantum technologies. We develop a noise-adaptive dynamical decoupling framework that combines analytical open-quantum-system modeling with machine-learning-based forecasting for a qubit interacting with random telegraph noise. Unlike conventional dynamical decoupling protocols based on fixed pulse schedules, the proposed approach continuously forecasts short-time coherence evolution and adaptively applies control pulses according to the instantaneous noise dynamics. We investigate stationary and non-stationary environments spanning both Markovian and non-Markovian regimes. Numerical simulations demonstrate that the machine-learning-assisted adaptive control strategy substantially outperforms conventional periodic dynamical decoupling while using a comparable number of control pulses. The improvement becomes particularly pronounced in non-Markovian and non-stationary regimes, where memory effects, coherence revivals, and temporally evolving noise strongly limit the effectiveness of static pulse protocols. These results establish predictive machine-learning-assisted dynamical decoupling as a promising and scalable framework for adaptive quantum control in realistic noisy quantum devices.

\end{abstract}
\maketitle
\section{Introduction}\label{sec1}

Quantum coherence lies at the heart of many emerging quantum technologies, including quantum computation \cite{nielsen2010quantum}, quantum communication \cite{Gisin2007}, and quantum sensing \cite{Degen2017}. The ability of quantum systems to exist in coherent superpositions enables powerful quantum algorithms \cite{shor1994,grover1997}, enhanced metrology \cite{Giovannetti2004}, and secure communication protocols \cite{Gisin2002} that outperform their classical counterparts. In practice, however, quantum systems are never perfectly isolated. Interactions with surrounding environmental degrees of freedom inevitably disturb the system and lead to the gradual loss of quantum coherence. This process, known as decoherence, remains one of the principal obstacles to the realization of scalable quantum technologies \cite{breuer,Zurek2003,Schlosshauer2005,rivas2}.

The theory of open quantum systems provides the fundamental framework for understanding how environmental interactions influence quantum dynamics \cite{breuer,rivas,rivas2,banerjee2018open}. By tracing over environmental degrees of freedom, the system evolves according to effective non-unitary dynamics that lead to decoherence and dissipation. Among the various decoherence mechanisms, pure dephasing plays a particularly important role in many solid-state quantum platforms \cite{breuer2002theory,Schlosshauer2005,Paladino2014}. In a dephasing process, environmental fluctuations randomize the relative phase of quantum superpositions while largely preserving the population probabilities. Since phase coherence is essential for quantum interference, quantum information processing, and coherent control, suppressing dephasing noise remains a central challenge in quantum-device engineering.

Several approaches have been developed to mitigate the detrimental effects of decoherence and preserve quantum coherence during system evolution. One prominent strategy is quantum error correction, where logical quantum information is encoded into larger Hilbert spaces that enable errors to be detected and corrected \cite{shor,Steane1996,Gottesman1997}. Another widely used approach is dynamical decoupling, in which sequences of control pulses are applied to suppress environmental interactions and prolong coherence times \cite{Viola1998,Viola1999,Khodjasteh2005,Uhrig2007}. The basic principle of dynamical decoupling is closely related to the spin-echo technique developed in nuclear magnetic resonance \cite{Hahn1950}, where rapid control operations refocus phase accumulation induced by environmental fluctuations. Motivated by this idea, a variety of dynamical decoupling protocols have been developed, including periodic dynamical decoupling \cite{Viola1998,Viola1999}, concatenated dynamical decoupling \cite{Khodjasteh2005}, and Uhrig dynamical decoupling \cite{Uhrig2007}. Although these protocols differ in pulse timing and control structure, they share the common objective of suppressing decoherence through carefully designed pulse sequences \cite{Viola1999,Suter2016}.

Despite their success, the effectiveness of dynamical decoupling protocols depends strongly on the statistical and spectral properties of the surrounding environment. Pulse sequences optimized for stationary noise can become inefficient when environmental characteristics evolve in time or when the noise spectrum is only partially known \cite{Viola1999, Khodjasteh2005,Biercuk2009,Suter2016}. Consequently, fixed pulse schedules may fail to provide optimal coherence protection in realistic noisy quantum devices.

In many solid-state quantum platforms, the environment contains microscopic fluctuators such as charge traps, defects, or impurities that randomly switch between metastable configurations \cite{Machlup1954,Paladino2002,Galperin2006,Paladino2014}. These bistable fluctuators generate stochastic fluctuations in the qubit energy splitting, producing a form of dephasing noise known as random telegraph noise (RTN). RTN has been widely observed in superconducting circuits and semiconductor quantum devices and constitutes an important source of decoherence in solid-state qubits \cite{Ithier2005,Paladino2014,banerjee2018open, PhysRevA.99.042128,kumar,Utagi2020Temporal}. Depending on the fluctuator switching rate and system--environment coupling strength, RTN can induce both Markovian and non-Markovian quantum dynamics \cite{breuer2002theory,rivas,rivas2}. In realistic experimental settings, however, the parameters governing the fluctuators may themselves evolve slowly in time due to environmental drift, device instabilities, or interactions between defects. Such effects generate effectively non-stationary noise processes \cite{Ithier2005,Paladino2014,Faoro2015}, for which fixed dynamical decoupling schedules may no longer remain optimal.

The limitations of static pulse protocols in non-stationary environments have motivated increasing interest in adaptive and learning-based approaches to quantum control. In recent years, machine-learning techniques have emerged as powerful tools for optimizing quantum control protocols and tailoring dynamical decoupling sequences to experimentally relevant noise environments \cite{Bukov2018,Carleo2019,Fosel2018,Niu2019}. Learning-based control strategies have demonstrated enhanced noise suppression by automatically identifying pulse sequences adapted to the underlying quantum device \cite{Rahman2024}. Predictive approaches have also been employed to estimate the future evolution of qubit observables under stochastic noise, enabling real-time mitigation of decoherence effects \cite{Mavadia2017}. More recently, anticipatory frameworks exploiting temporal correlations in environmental dynamics have been proposed to forecast decoherence processes before they occur, allowing control operations to be applied preemptively \cite{Maan2026}. These developments highlight the growing role of predictive and adaptive techniques in the control of open quantum systems.

Motivated by these advances, it is natural to ask whether the short-time evolution of a quantum system can be predicted and used to dynamically guide the application of control pulses. For a qubit undergoing pure dephasing, the system coherence is directly related to the off-diagonal elements of the density matrix and can be experimentally accessed through transverse observables \cite{nielsen2010quantum,breuer}. Monitoring the evolution of these observables therefore provides direct information about the decoherence dynamics and offers a natural route toward predictive quantum control.

In this work, we develop a noise-adaptive predictive dynamical decoupling framework for a qubit interacting with an RTN environment. We first derive an analytical description of the coherence dynamics of the observable $
X(t)=\langle \sigma_x(t)\rangle$, leading to a closed differential equation governing the short-time evolution of the system coherence. This analytical model enables short-time forecasting of the coherence evolution and provides a physics-based predictor for adaptive pulse control.

In parallel, a machine-learning model is trained on time-series data of measured qubit observables in order to learn the underlying system dynamics and predict the future evolution of coherence directly from observed time traces. The analytical and data-driven predictors are then used to determine when dynamical decoupling pulses should be applied.

Unlike conventional periodic dynamical decoupling protocols, where pulse timing is fixed in advance, the proposed approach applies control pulses adaptively whenever the predicted dynamics indicate an imminent loss of coherence. The resulting framework therefore implements a noise-adaptive control strategy capable of responding dynamically to time-dependent environmental fluctuations.

We investigate this predictive framework for both stationary and non-stationary random telegraph noise environments spanning Markovian and non-Markovian regimes. Non-stationarity is introduced by allowing the switching rate $\kappa(t)$ and coupling strength $v(t)$ of the noise process to vary slowly in time, thereby modeling environmental drift and temporally evolving fluctuations. Numerical simulations demonstrate that the predictive strategy significantly improves coherence preservation compared with conventional periodic dynamical decoupling, particularly in non-Markovian and non-stationary environments where fixed pulse schedules become inefficient.

The remainder of this paper is organized as follows. 
In Sec.~\ref{sec2}, we introduce the theoretical framework and derive the open quantum system model describing a qubit interacting with random telegraph noise. We also develop the analytical short-time predictor and introduce the stationary and non-stationary RTN models used throughout this work.  In Sec.~\ref{sec:ML}, we present the machine-learning framework used to predict the future coherence dynamics from time-series measurements.   In Sec.~\ref{sec:DD}, we introduce the dynamical decoupling protocols considered in this work, including both conventional periodic dynamical decoupling and the proposed machine-learning-assisted adaptive control strategy.  In Sec.~\ref{sec:results}, we present the numerical results and compare the performance of the predictive dynamical decoupling framework with conventional pulse protocols in both stationary and non-stationary Markovian and non-Markovian environments.  Finally, Sec.~\ref{sec:conclusion} summarizes the main results and discusses possible future directions for predictive quantum control in realistic noisy quantum devices.

%%%%%%%%%%%%%%%%%%%%%%%%%%%%%%%%%%%%%%%%%%%

\section{Theoretical Framework and Open Quantum System Model}
\label{sec2}

In this section we develop the theoretical framework
describing a qubit interacting with a dephasing
environment. Starting from a microscopic
system-environment Hamiltonian, we derive the reduced
open-system dynamics of the qubit and obtain the
equations governing the evolution of experimentally
accessible observables. Expressing the dynamics in terms
of Bloch observables leads to a closed equation for the
coherence of the system, which forms the basis of the
predictive control framework developed in this work.

We then specialize the general formalism to RTN, a paradigmatic model of
decoherence in solid-state quantum devices. Both
stationary and non-stationary RTN models are considered
in order to describe environments whose statistical
properties evolve in time, thereby providing a realistic
framework for investigating adaptive dynamical
decoupling in Markovian and non-Markovian regimes.

\subsection{Microscopic qubit--environment Hamiltonian} We consider a single qubit interacting with an external environment. The qubit is described within a two-dimensional Hilbert space $\mathcal{H}_S=\mathrm{span}\{|0\rangle,|1\rangle\}$, where $|0\rangle$ and $|1\rangle$ denote the computational basis states of the system. The environment is modeled as a large bosonic reservoir with Hilbert space $\mathcal{H}_B$, consisting of a collection of independent harmonic oscillator modes. The total Hilbert space of the combined system-environment configuration is therefore $\mathcal{H}=\mathcal{H}_S\otimes\mathcal{H}_B$. 
The total Hamiltonian governing the joint dynamics of the qubit and the environment is written as \begin{equation} H = H_S + H_B + H_{SB}, \end{equation} where $H_S$, $H_B$, and $H_{SB}$ denote the system, environment, and interaction Hamiltonians, respectively. The qubit Hamiltonian is given by \begin{equation} H_S = \frac{\omega_0}{2}\sigma_z , \end{equation} where $\omega_0$ is the qubit transition frequency and $\sigma_z$ is the Pauli-$z$ operator. The environment is modeled as a bosonic bath composed of independent harmonic oscillator modes, \begin{equation} H_B = \sum_k \omega_k b_k^\dagger b_k , \end{equation} where $\omega_k$ is the frequency of the $k$th mode, and $b_k^\dagger$ and $b_k$ are bosonic creation and annihilation operators satisfying the commutation relation $ [b_k,b_{k'}^\dagger]=\delta_{kk'}$.  The interaction between the qubit and the environment is described by \begin{equation} H_{SB} = \sigma_z\sum_k g_k(b_k+b_k^\dagger), \label{HSB} \end{equation} where $g_k$ denotes the coupling strength between the qubit and the $k$th bath mode. The bath operators $b_k^\dagger$ and $b_k$ describe the creation and annihilation of environmental excitations, while the operator $\sigma_z$ acts on the qubit degree of freedom. Because the interaction Hamiltonian is proportional to $\sigma_z$, it commutes with the system Hamiltonian, $ [H_S,H_{SB}]=0$.  Consequently, the interaction does not induce transitions between the qubit energy eigenstates $|0\rangle$ and $|1\rangle$, and therefore the population probabilities remain unchanged during the evolution. Instead, the environment continuously modulates the relative phase accumulated between the two energy levels. The resulting dynamics therefore correspond to pure dephasing, in which quantum coherence decays without energy exchange between the qubit and the environment. 

Having specified the microscopic system--environment
interaction, we now derive the effective open-system
dynamics of the qubit. The full state of the combined
system and environment is described by the total density
matrix $\rho_{\mathrm{tot}}(t)$, whose evolution is
governed by the total Hamiltonian $H$. Since the
combined system--environment configuration forms a
closed quantum system, the total evolution is unitary.

In realistic experimental situations, however, the
environmental degrees of freedom are not directly
accessible. The observable dynamics of the qubit are
therefore described by the reduced density matrix
\begin{equation}
\rho_S(t)=\mathrm{Tr}_B[\rho_{\mathrm{tot}}(t)],
\end{equation}
obtained by tracing over the environmental Hilbert
space.

Under the standard Born-Markov assumptions of open
quantum system theory, namely weak
system-environment coupling, an initially factorized
system-bath state, and rapidly decaying bath
correlations
\cite{breuer2002theory,Weiss,banerjee2018open},
the reduced dynamics of the qubit can be described by a
time-local master equation in
Gorini-Kossakowski-Sudarshan-Lindblad  (GKSL)~\cite{gorini,lindblad} form.

The reduced dynamics can therefore be written as
\begin{equation}
\frac{d\rho_S(t)}{dt}
=
-i[H_S,\rho_S(t)]
+
\gamma(t)
\left(
L\rho_S(t)L^\dagger
-\frac{1}{2}
\{L^\dagger L,\rho_S(t)\}
\right),
\label{lindblad_general}
\end{equation}
where $L$ is the Lindblad operator associated with the
system-environment interaction and $\gamma(t)$ is the
time-dependent dephasing rate.

For the interaction Hamiltonian in
Eq.~(\ref{HSB}), the coupling operator is proportional
to $\sigma_z$, yielding $
L=\sigma_z $.

Substituting this operator into
Eq.~(\ref{lindblad_general}) gives
\begin{equation}
\frac{d\rho_S(t)}{dt}
=
-i\left[
\frac{\omega_0}{2}\sigma_z,\rho_S(t)
\right]
+
\gamma(t)
\left(
\sigma_z\rho_S(t)\sigma_z-\rho_S(t)
\right),
\label{master}
\end{equation}
which describes pure dephasing of the qubit.

The coefficient $\gamma(t)$ characterizes the influence
of the environment on the coherence dynamics of the
system. In Markovian environments $\gamma(t)$ remains
positive and typically constant, leading to monotonic
decoherence. In contrast, non-Markovian environments
may exhibit time-dependent decay rates that become
temporarily negative, reflecting environmental memory
effects and information backflow from the environment
to the qubit.

\subsection{Bloch dynamics and analytical short-time predictor}

To describe experimentally measurable quantities, we
express the reduced qubit dynamics in terms of the
Bloch observables
\begin{equation} \begin{aligned} X(t)&=\Tr[\rho_S(t)\sigma_x],\\ Y(t)&=\Tr[\rho_S(t)\sigma_y],\\ Z(t)&=\Tr[\rho_S(t)\sigma_z]. \end{aligned} \end{equation}

which correspond to the Cartesian components of the
Bloch vector.

Substituting the master equation
(\ref{master}) into the time derivatives of these
observables and using the Pauli commutation relations
yields the coupled equations
\begin{align}
\dot X(t) &= -\omega_0Y(t)-2\gamma(t)X(t), \\
\dot Y(t) &= \omega_0X(t)-2\gamma(t)Y(t),
\label{eq:Y} \\
\dot Z(t) &= 0 .
\end{align}

These equations show that pure dephasing does not alter
the population difference $Z(t)$, while the transverse
coherence components $X(t)$ and $Y(t)$ decay due to the
interaction with the environment.

To obtain a closed dynamical equation for the coherence
observable $X(t)$, we differentiate the first Bloch
equation with respect to time and eliminate $Y(t)$
using Eq.~(\ref{eq:Y}). This yields
\begin{equation}
\ddot X(t)
+
4\gamma(t)\dot X(t)
+
\left[
\omega_0^2
+
4\gamma^2(t)
+
2\dot\gamma(t)
\right]X(t)
=0 ,
\label{general_predictive_equation}
\end{equation}
which describes the coherence dynamics as a
non-stationary damped oscillator with time-dependent
coefficients.

Since Eq.~(\ref{general_predictive_equation}) provides
a closed equation governing the short-time coherence
dynamics, it naturally enables the construction of a
local predictive model for the future evolution of the
system. For a sufficiently small time step $\Delta t$,
the coherence observable can be expanded about the
current time $t$ using a Taylor expansion
\cite{SarkkaSolin2019},

\begin{equation}
X(t+\Delta t)
=
X(t)
+
\Delta t\,\dot X(t)
+
\frac{\Delta t^2}{2}\ddot X(t)
+
\mathcal O(\Delta t^3).
\label{eq:taylor}
\end{equation}

Substituting
Eq.~(\ref{general_predictive_equation})
for $\ddot X(t)$ gives

\begin{align}
X(t+\Delta t)
&=
X(t)
+
\Delta t\,\dot X(t)
\nonumber\\
&
-
\frac{\Delta t^2}{2}
\Big[
4\gamma(t)\dot X(t)
+
\big(
\omega_0^2
+
4\gamma^2(t)
+
2\dot\gamma(t)
\big)X(t)
\Big].
\label{analytic_predictor}
\end{align}

Equation~(\ref{analytic_predictor}) therefore provides
an analytical short-time predictor for the coherence
observable. Using the instantaneous values of
$X(t)$ and $\dot X(t)$, the expression enables the
prediction of the coherence at a slightly later time
$t+\Delta t$, forming the basis of the predictive
dynamical decoupling framework developed in this work.

%%%%%%%%%%%%%%%%%%%%%%%%%%%%%%%%%%

\subsection{Random Telegraph Noise}
\label{subsec:rtn}

Random telegraph noise (RTN) is one of the standard models of
dephasing noise in solid-state quantum devices, where the qubit
frequency is modulated by bistable environmental defects that switch
randomly between two configurations
\cite{Paladino2014,PhysRevA.99.042128,NaikooBanerjee2020}.
Examples include charge traps, fluctuating impurities, and
two-level defects in superconducting or semiconductor platforms.
In this model the qubit is subjected to a classical stochastic field
that changes sign at random times, thereby producing phase noise.

A convenient description of qubit dephasing under
random telegraph noise is obtained from the stochastic
Hamiltonian
\begin{equation}
H(t)=\frac{\omega_0}{2}\sigma_z+\xi(t)\sigma_z,
\label{eq:rtn_hamiltonian}
\end{equation}
where $\xi(t)$ represents a stochastic telegraph process
that randomly switches between two discrete values
\cite{Paladino2014,banerjee2018open,
PhysRevA.99.042128}, where $
\xi(t)=v\,\chi(t)$, $v$ is the coupling strength between the qubit and the fluctuator,
and $\chi(t)$ is a dichotomic stochastic process taking the values
$\pm 1$.
Thus the fluctuating field switches between the two values $\pm v$.

For stationary symmetric RTN, the process $\chi(t)$ is assumed to
switch from $+1$ to $-1$ and from $-1$ to $+1$ with the same rate
$\kappa$.
In this convention the stationary autocorrelation function is
\begin{equation}
\langle \chi(t)\chi(s)\rangle = e^{-2\kappa|t-s|},
\end{equation}
and therefore the noise autocorrelation becomes
\begin{equation}
\langle \xi(t)\xi(s)\rangle = v^2 e^{-2\kappa|t-s|}.
\label{eq:rtn_corr_consistent}
\end{equation}
The corresponding bath correlation time is
\begin{equation}
\tau_B=\frac{1}{2\kappa}.
\end{equation}

Because the Hamiltonian in Eq.~(\ref{eq:rtn_hamiltonian}) is diagonal
in the $\sigma_z$ basis, the RTN field does not induce transitions
between the qubit energy eigenstates and acts purely as a dephasing
mechanism.
For a single realization of the noise, the off-diagonal element of the
density matrix acquires the stochastic phase factor
\begin{equation}
\rho_{01}(t)=e^{-i\omega_0 t}\,
e^{-2i\int_0^t \xi(\tau)\,d\tau}\,
\rho_{01}(0).
\end{equation}
Averaging over the stochastic process defines the decoherence function
\begin{equation}
\Lambda(t)=\left\langle
e^{-2i\int_0^t \xi(\tau)\,d\tau}
\right\rangle ,
\label{eq:lambda_def}
\end{equation}
so that the qubit coherence evolves as
\begin{equation}
X(t)=\Lambda(t)X(0).
\label{eq:x_lambda}
\end{equation}

To derive a closed equation for $\Lambda(t)$, it is useful to introduce
the conditional averages
\begin{equation}
\Lambda_\pm(t)=
\left\langle
e^{-2iv\int_0^t \chi(\tau)\,d\tau}
\right\rangle_{\chi(0)=\pm1},
\end{equation}
corresponding to trajectories that begin in the states $\chi(0)=+1$
and $\chi(0)=-1$, respectively.
During an infinitesimal interval $dt$, the fluctuator either remains in
its current state with probability $1-\kappa dt$ or switches to the
opposite state with probability $\kappa dt$.
Keeping only terms to first order in $dt$ therefore gives
\begin{align}
\Lambda_+(t+dt)
&=(1-\kappa dt)\,e^{-2iv\,dt}\Lambda_+(t)
+\kappa dt\,e^{-2iv\,dt}\Lambda_-(t), \\
\Lambda_-(t+dt)
&=(1-\kappa dt)\,e^{+2iv\,dt}\Lambda_-(t)
+\kappa dt\,e^{+2iv\,dt}\Lambda_+(t).
\end{align}
Expanding to first order in $dt$ and rearranging yields the coupled
equations
\begin{align}
\dot{\Lambda}_+(t)
&= -2iv\,\Lambda_+(t)-\kappa \Lambda_+(t)+\kappa \Lambda_-(t), \\
\dot{\Lambda}_-(t)
&= +2iv\,\Lambda_-(t)-\kappa \Lambda_-(t)+\kappa \Lambda_+(t).
\end{align}

We now introduce the symmetric and antisymmetric combinations
\begin{equation}
\Lambda(t)=\frac{\Lambda_+(t)+\Lambda_-(t)}{2},
\qquad
\Delta(t)=\frac{\Lambda_+(t)-\Lambda_-(t)}{2}.
\end{equation}
Adding and subtracting the two equations above gives
\begin{align}
\dot{\Lambda}(t) &= -2iv\,\Delta(t), \\
\dot{\Delta}(t) &= -2iv\,\Lambda(t)-2\kappa\,\Delta(t).
\end{align}
Differentiating the first equation and eliminating $\Delta(t)$ using
the second yields
\begin{equation}
\ddot{\Lambda}(t)+2\kappa\dot{\Lambda}(t)+4v^2\Lambda(t)=0.
\label{eq:lambda_rtn_ode}
\end{equation}

Since $X(t)=X(0)\Lambda(t)$, the same equation holds for the coherence
observable itself:
\begin{equation}
\ddot X(t)+2\kappa \dot X(t)+4v^2 X(t)=0.
\label{eq:rtn_equation}
\end{equation}
This is the RTN-specific coherence equation used in the present work.
It has the form of a damped oscillator, where $\kappa$ plays the role
of an effective damping parameter and $v$ sets the oscillation scale.

The initial conditions follow directly from
$\Lambda(0)=1$ and $\dot{\Lambda}(0)=0$, yielding the exact solution
\begin{equation}
\Lambda(t)
=
e^{-\kappa t}
\left[
\cos(\mu\kappa t)+\frac{\sin(\mu\kappa t)}{\mu}
\right],
\label{eq:lambda_rtn}
\end{equation}
with
\begin{equation}
\mu=\sqrt{\left(\frac{2v}{\kappa}\right)^2-1}.
\label{eq:mu_rtn}
\end{equation}

\begin{figure}[t]
    \centering
    \includegraphics[width=1.0 \linewidth]{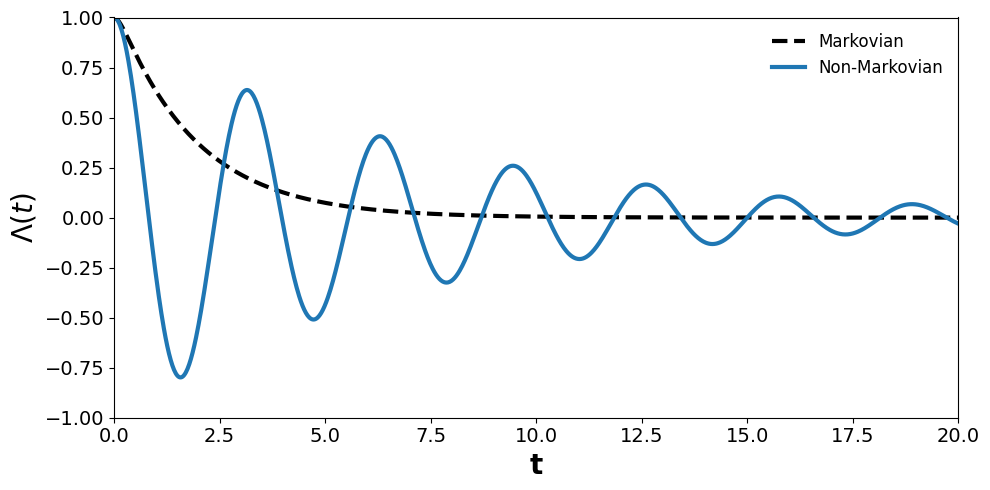}
    \caption{Decoherence function $\Lambda(t)$ for a qubit subject to
    RTN. The dashed black curve corresponds to the
    Markovian regime, where $2v<\kappa$ and the coherence decays
    monotonically. The solid blue curve corresponds to the
    non-Markovian regime, where $2v>\kappa$ and the coherence exhibits
    damped oscillations and revivals. The latter behavior reflects the
    presence of memory effects and information backflow in the effective
    dephasing dynamics.}
    \label{fig:rtn_markov_nonmarkov}
\end{figure}

The qualitative character of the dynamics is determined by the ratio
$2v/\kappa$.
When $2v<\kappa$, the parameter $\mu$ becomes imaginary and the
coherence decays monotonically, corresponding to the Markovian regime.
By contrast, when $2v>\kappa$, $\mu$ is real and the coherence
exhibits damped oscillations and revivals.
These revivals are a manifestation of environmental memory and are
therefore associated with non-Markovian behavior.
This transition between the two regimes is illustrated in
Fig.~\ref{fig:rtn_markov_nonmarkov}, where the dashed black curve
shows the monotonic decay characteristic of the Markovian regime,
while the solid blue curve displays the damped oscillatory behavior
that arises in the non-Markovian regime.
\cite{Tiwari2025}.

In realistic situations the characteristics of the fluctuator may drift
with time, for example because of temperature changes, device bias
instabilities, or slow variations in the surrounding environment.
A simple phenomenological extension is therefore obtained by allowing
the effective coupling strength and switching rate to become time
dependent,
\begin{equation}
v \rightarrow v(t), \qquad
\kappa \rightarrow \kappa(t).
\end{equation}
Replacing the constant parameters in Eq.~(\ref{eq:rtn_equation}) then
gives the generalized non-stationary RTN equation
\begin{equation}
\ddot X(t)
+
2\kappa(t)\dot X(t)
+
4v^2(t)X(t)
=0 ,
\label{eq:nonstationary_rtn}
\end{equation}
which forms the starting point for the predictive framework developed
in the following sections.

A microscopic route to RTN dynamics has recently been discussed in a
central-spin setting with random stochastic system-bath coupling,
providing further support for using RTN as an effective dephasing model
beyond a purely phenomenological description.~\cite{Tiwari2025}
%%%%%%%%%%%%%%%%%%%%%%%%%%%%%%%%%%%%%

To describe such situations we extend the RTN model by allowing the
coupling strength $v$ and the switching rate of the fluctuator $\kappa$ to vary in time. This generalization is physically motivated by two complementary ideas.

First, the fluctuator switching statistics may become
time dependent, as in random telegraph signal models
with time-dependent capture and emission probabilities
under cyclo-stationary excitation
\cite{Machlup1954,Paladino2014,Faoro2015}.
Second, the effective coupling between the system and
its environment may itself vary in time, as commonly
encountered in open quantum system models with
time-dependent system--bath interaction strengths
\cite{BreuerPetruccione,Weiss,banerjee2018open}.

\subsubsection{Two models of non-stationary RTN}

To investigate the robustness of the predictive framework
against different forms of non-stationarity, we consider
two representative models for the time dependence of the
RTN parameters \(v(t)\) and \(\kappa(t)\).

\paragraph{Model 1: Gaussian environmental drift}

We first introduce a phenomenological model describing
localized environmental drift. In realistic solid-state
quantum devices, the parameters governing microscopic
fluctuators may evolve slowly in time due to temperature
variations, bias fluctuations, or the activation and
deactivation of nearby defects
\cite{Machlup1954,Paladino2002,Bergli2009,
Paladino2014,Faoro2015,Galperin2006}. Such slow
variations constitute an important source of
non-stationary noise in qubit platforms.

To model this behavior, we introduce a slow drift
variable \(q(t)\) that becomes significant only within
a finite temporal interval. A convenient smooth
representation is provided by a Gaussian envelope
commonly used in noise modeling and control theory to
describe localized temporal perturbations
\cite{AllenEberly1987,SarkkaSolin2019},
\begin{equation}
q(t)=\exp\!\left[-\frac{(t-t_0)^2}{2\sigma^2}\right],
\label{eq:q_gaussian}
\end{equation}

where \(t_0\) denotes the time at which the
environmental perturbation reaches its maximum effect,
and \(\sigma\) determines the temporal width of the
Gaussian envelope. Larger values of \(\sigma\)
correspond to slowly varying environmental drift spread
over longer durations, whereas smaller values describe
more localized perturbations. 

Assuming that the RTN parameters respond linearly to the
drift variable gives
\begin{equation}
v(t)=v_0\bigl[1+A_v q_v(t)\bigr],
\qquad
\kappa(t)=\kappa_0\bigl[1+A_\kappa q_\kappa(t)\bigr],
\label{eq:linear_drift_response}
\end{equation}
where \(v_0\) and \(\kappa_0\) denote the baseline
coupling strength and switching rate, respectively, while
\(A_v\) and \(A_\kappa\) determine the magnitude of the
drift.
Assuming Gaussian drift envelopes for the environmental
modulation yields the non-stationary RTN model
\begin{equation}\label{eq:vt}
v(t)=v_0\left[1+A_v
\exp\!\left(-\frac{(t-t_v)^2}{2\sigma_v^2}\right)\right],
\end{equation}
\begin{equation}\label{eq:kappat}
\kappa(t)=\kappa_0\left[1+A_\kappa
\exp\!\left(-\frac{(t-t_\kappa)^2}{2\sigma_\kappa^2}\right)\right].
\end{equation}

The parameters appearing in Eqs. (\ref{eq:vt}) and (\ref{eq:kappat}) have the following physical interpretation. The quantities ($v_0$) and ($\kappa_0$) denote the baseline qubit--environment coupling strength and the baseline fluctuator switching rate, respectively. The dimensionless coefficients ($A_v$) and ($A_\kappa$) determine the relative amplitude of the environmental drift and quantify the strength of the non-stationary modulation. The parameters ($t_v$) and ($t_\kappa$) specify the temporal locations at which the Gaussian perturbations reach their maximum values, while ($\sigma_v$) and ($\sigma_\kappa$) characterize the temporal widths of the Gaussian envelopes and therefore determine how slowly or rapidly the environmental drift develops in time.

In the numerical simulations presented in this work, the stationary Markovian regime was generated using parameters satisfying ($2v_0 < \kappa_0$), while the non-Markovian regime was obtained using ($2v_0 > \kappa_0$). Unless otherwise stated, the simulations employed dimensionless units with ($\omega_0 = 1$). Representative parameters used throughout the simulations were chosen within the ranges $
v_0 \sim 0.2 - 1.2$, $\kappa_0 \sim 0.5 - 3.0$, 
with drift amplitudes $
A_v, A_\kappa \sim 0.1 - 0.5$, 
and Gaussian widths $
\sigma_v, \sigma_\kappa \sim 5 - 20$.
The pulse-control and machine-learning simulations used identical environmental realizations in order to allow a direct comparison between conventional periodic dynamical decoupling and the predictive DD+ML framework. Substituting Eqs.~(\ref{eq:vt}) and
(\ref{eq:kappat}) into the RTN coherence equation, Eq.~\ref{eq:rtn_equation}, gives
\begin{align}\label{eq:rtn_gaussian_model}
\ddot X(t)
+
2\kappa_0\left[1+A_\kappa
\exp\!\left(-\frac{(t-t_\kappa)^2}{2\sigma_\kappa^2}\right)\right]\dot X(t)
\nonumber\\
+
4v_0^2\left[1+A_v
\exp\!\left(-\frac{(t-t_v)^2}{2\sigma_v^2}\right)\right]^2 X(t)
=0.
\end{align}

Equation~(\ref{eq:rtn_gaussian_model}) describes a
non-stationary damped oscillator with localized temporal
modulation of both the effective damping and oscillation
scale. The model therefore captures situations in which
the spectral properties of the environment evolve slowly
during the system dynamics
\cite{Paladino2014,Faoro2015}.

To construct an analytical short-time predictor for the
coherence dynamics, we expand the observable \(X(t)\)
around time \(t\) using a Taylor expansion, Eq.~\ref{eq:taylor}. Substituting Eq.~(\ref{eq:rtn_gaussian_model}) for
\(\ddot X(t)\) yields the Gaussian-drift predictive
model

\begin{align}
X(t+\Delta t)
&=
X(t)
+
\Delta t\,\dot X(t)
\nonumber\\
&\quad
-
\frac{\Delta t^2}{2}
\Bigg[
2\kappa_0
\left(
1+A_\kappa
\exp\!\left(
-\frac{(t-t_\kappa)^2}{2\sigma_\kappa^2}
\right)
\right)\dot X(t)
\nonumber\\
&\qquad\qquad
+
4v_0^2
\left(
1+A_v
\exp\!\left(
-\frac{(t-t_v)^2}{2\sigma_v^2}
\right)
\right)^2
X(t)
\Bigg].
\label{eq:rtn_gaussian_predictor}
\end{align}

Equation~(\ref{eq:rtn_gaussian_predictor}) provides the
analytical predictor.
\paragraph{Model 2: Oscillatory environmental modulation}

While Model~1 describes localized environmental drift,
realistic solid-state environments may also exhibit
quasi-periodic fluctuations arising from interactions
among fluctuators, slow thermal cycles, or externally
driven perturbations
\cite{Machlup1954,Galperin2006,Bergli2009,
Paladino2014}. To model such behavior, we consider a
second non-stationary RTN model in which both the
switching dynamics and coupling strength vary
continuously in time.

We begin with a general two-state telegraph process in
which transitions between the fluctuator states occur
with time-dependent rates
$
0 \xrightarrow{k(t)} 1
$
and
$
1 \xrightarrow{l(t)} 0.
$
The conditional probability
\(p_{1,1}^{(t)}(s)\)
satisfies
\begin{equation}
\frac{d}{ds}p_{1,1}^{(t)}(s)
=
-\bigl[k(s)+l(s)\bigr]p_{1,1}^{(t)}(s)+l(s),
\label{eq:general_telegraph_eq}
\end{equation}
which represents the time-dependent extension of the
classical telegraph process introduced by Machlup
\cite{Machlup1954}.

Following the cyclo-stationary model investigated by
da Silva \textit{et al.}
\cite{daSilva2009},
we assume sinusoidally modulated transition rates,
\begin{equation}
k(t)=c_1\bigl[2-\sin(\Omega t+\phi_1)\bigr],
\quad
l(t)=c_2\bigl[2+\sin(\Omega t+\phi_2)\bigr],
\label{eq:dlw_kl_general}
\end{equation}
where \(c_1\) and \(c_2\) determine the baseline
transition rates, \(\Omega\) is the modulation
frequency, and \(\phi_1,\phi_2\) are phase shifts.

The effective RTN switching rate is defined as
\begin{equation}
\kappa(t)=\frac{k(t)+l(t)}{2},
\label{eq:kappa_eff_def}
\end{equation}
which yields
\begin{align}
\kappa(t)
&=
(c_1+c_2)
+
\frac{1}{2}
\left[
-c_1\sin(\Omega t+\phi_1)
+
c_2\sin(\Omega t+\phi_2)
\right].
\label{eq:kappa_intermediate}
\end{align}

More generally, periodically modulated switching
dynamics can be represented through a Fourier expansion.
Retaining only the first two harmonics, the effective
switching rate is modeled as
\begin{equation}
\kappa_{\mathrm{osc}}(t)
=
\kappa_0
+
\kappa_1\sin(\Omega t+\phi_{\kappa,1})
+
\kappa_2\sin(2\Omega t+\phi_{\kappa,2}),
\label{eq:kappa_model2}
\end{equation}
where \(\kappa_0\) denotes the average switching rate,
while \(\kappa_1\) and \(\kappa_2\) characterize the
strength of the oscillatory modulation.

We next consider the fluctuating field amplitude.
In open quantum systems, time-dependent system--bath
couplings naturally arise in structured environments,
driven reservoirs, and engineered noise processes
\cite{Kofman2000,Kofman2001,BreuerPetruccione}. Since
the RTN field is written as
$
\xi(t)=v(t)\chi(t),
$
the quantity \(v(t)\) plays the role of an effective
system--environment coupling strength.

Motivated by periodically modulated environments, we
similarly expand the coupling amplitude in a Fourier
series and retain the first two harmonics,
\begin{equation}
v_{\mathrm{osc}}(t)
=
v_0
+
v_1\sin(\Omega t+\phi_{v,1})
+
v_2\sin(2\Omega t+\phi_{v,2}),
\label{eq:v_model2}
\end{equation}
where \(v_0\) is the average coupling strength, while
\(v_1\) and \(v_2\) determine the amplitude of the
oscillatory modulation.

Substituting Eqs.~(\ref{eq:kappa_model2}) and
(\ref{eq:v_model2}) into the RTN coherence equation
yields
\begin{equation}
\ddot X(t)
+
2\kappa_{\mathrm{osc}}(t)\dot X(t)
+
4v_{\mathrm{osc}}^2(t)X(t)
=0.
\label{eq:rtn_oscillatory_model}
\end{equation}

Equation~(\ref{eq:rtn_oscillatory_model}) describes a
non-stationary RTN environment with persistent
oscillatory modulation of both the switching dynamics
and the coupling strength, leading to richer coherence
dynamics than the localized Gaussian drift model.

To construct the corresponding analytical short-time
predictor, we expand the coherence observable \(X(t)\)
around time \(t\) using the Taylor expansion in
Eq.~(\ref{eq:taylor}). Substituting
Eq.~(\ref{eq:rtn_oscillatory_model}) for
\(\ddot X(t)\) gives

\begin{equation}
\begin{aligned}
X(t+\Delta t)
&=
X(t)
+
\Delta t\,\dot X(t)
\\
&\quad
-
\frac{\Delta t^2}{2}
\Big[
2\kappa_{\mathrm{osc}}(t)\dot X(t)
+
4v_{\mathrm{osc}}^2(t)X(t)
\Big].
\end{aligned}
\label{eq:rtn_oscillatory_predictor}
\end{equation}

Equation~(\ref{eq:rtn_oscillatory_predictor}) therefore
provides the analytical predictor for the oscillatory
non-stationary RTN model.

\begin{figure*}[t]
\centering
\includegraphics[width=\linewidth]{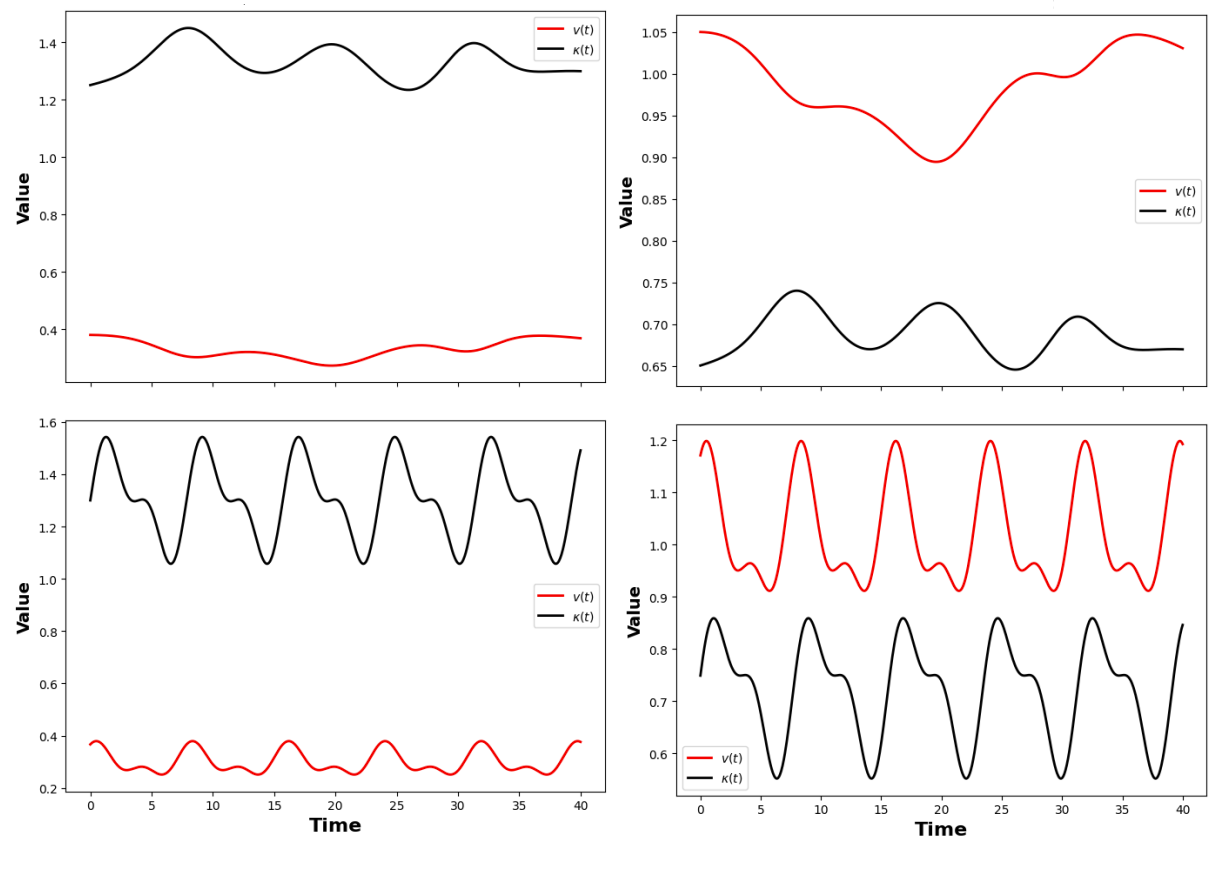}
\caption{
Time-dependent RTN parameters for the two
non-stationary noise models considered in this work.
The red curves denote the coupling strength \(v(t)\),
while the black curves represent the effective
switching rate \(\kappa(t)\).
Top panels: Model~1 with localized Gaussian-like drift
of the RTN parameters.
Bottom panels: Model~2 with persistent oscillatory
modulation of both \(v(t)\) and \(\kappa(t)\).
The left column corresponds to parameter regimes with
\(\kappa(t) > v(t)\), leading predominantly to
Markovian coherence dynamics, whereas the right column
corresponds to regimes where \(v(t)\) becomes
comparable to or larger than \(\kappa(t)\), allowing
non-Markovian memory effects and coherence revivals to
emerge.
}
\label{fig:kv_models}
\end{figure*}

To illustrate the two non-stationary RTN models used
in this work, Fig.~\ref{fig:kv_models} shows the
time-dependent coupling strength \(v(t)\) and effective
switching rate \(\kappa(t)\) for representative
Markovian and non-Markovian parameter regimes. The
upper panels correspond to Model~1, characterized by
slow Gaussian-like environmental drift of the RTN
parameters, while the lower panels correspond to
Model~2, which exhibits persistent oscillatory
modulation of both \(v(t)\) and \(\kappa(t)\). The
left column represents the predominantly Markovian
regime, where the switching rate satisfies
\(\kappa(t) > v(t)\), leading to monotonic coherence
decay. In contrast, the right column corresponds to the
non-Markovian regime, where \(v(t)\) becomes comparable
to or larger than \(\kappa(t)\), allowing coherence
revivals and memory effects to emerge.
%%%%%%%%%%%%%%%%%%%%%%%%%%%%%%%%%%%%%%

%%%%%%%%%%%%%%%%%%%%%%%%%%%%%%%%%%%%%%%%%%%%%%%%
\section{Machine Learning Prediction of the Dynamics}
\label{sec:ML}

In the previous section we derived an analytical predictor for the
short–time evolution of the coherence observable
\(X(t)=\langle\sigma_x(t)\rangle\).
We now introduce a second method for forecasting the system dynamics
based on supervised machine learning.

Unlike the analytical approach, which relies on explicit knowledge of
the noise model, the machine learning predictor learns the dynamical
behavior directly from time–series data generated from numerical
simulations of the open quantum system.

The goal of the model is therefore to estimate the future value of the
coherence observable using a short history of previously measured
values.

\subsection{Construction of the Training Dataset}

The dataset used to train the machine learning model is generated from
numerical simulations of the open quantum system dynamics.
In particular, we compute the time evolution of the qubit interacting
with a random telegraph noise (RTN) environment using the QuTiP
framework \cite{qutip1,qutip2}.

From the simulated density matrix we extract the coherence observable

\begin{equation}
X(t)=\langle\sigma_x(t)\rangle
=
\text{Tr}[\rho(t)\sigma_x]
\end{equation}

at discrete time steps.

To construct the dataset for supervised learning we employ a sliding
window approach.

At each time step \(t_i\), the input feature vector is formed from the
previous five measurements of the observable

\begin{equation}
\mathbf{x}_i=
\big[
X(t_{i-5}),
X(t_{i-4}),
X(t_{i-3}),
X(t_{i-2}),
X(t_{i-1})
\big].
\end{equation}

The corresponding target label is the value of the observable at the
next time step

\begin{equation}
y_i = X(t_i).
\end{equation}

Each training example therefore consists of an input–output pair
\((\mathbf{x}_i,y_i)\), where the model receives a short temporal
history of the system and attempts to predict its immediate future
value.

The overall structure of the dataset is summarized in Table~\ref{tab:dataset}.

\begin{table}[h!]
\centering
\small
\begin{tabular}{cc}
\toprule
\textbf{Input Window $\mathbf{x}_i$} & \textbf{Target Output $y_i$} \\
\midrule
$[X(t_0),\ldots,X(t_4)]$ & $X(t_5)$ \\
$[X(t_1),\ldots,X(t_5)]$ & $X(t_6)$ \\
$[X(t_2),\ldots,X(t_6)]$ & $X(t_7)$ \\
$\vdots$ & $\vdots$ \\
\bottomrule
\end{tabular}
\caption{Structure of the supervised learning dataset generated using a sliding time window. 
Each input consists of the five most recent observations of the coherence observable, while the target output is the value at the next time step.}
\label{tab:dataset}
\end{table}
\begin{figure}[h]
\centering
\includegraphics[width=1.0\linewidth]{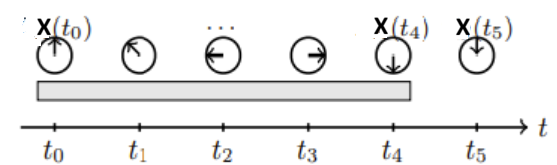}
\caption{Sliding-window construction of the training dataset.
At time \(t_i\) the model receives the five previous measurements
\([X(t_{i-5}),\ldots,X(t_{i-1})]\) and predicts the next value
\(X(t_i)\).}
\label{fig:window}
\end{figure}
\subsection{Neural Network Architecture}

The predictive model is implemented as a fully connected feedforward
neural network.

The input layer receives the five values of the sliding window

\[
[X(t_{i-5}),X(t_{i-4}),X(t_{i-3}),X(t_{i-2}),X(t_{i-1})].
\]

The network contains two hidden layers with 32 and 16 neurons,
respectively.

Each neuron in the first hidden layer computes

\begin{equation}
h_j^{(1)}
=
\mathrm{ReLU}
\left(
\sum_{k=1}^{5} w_{jk}^{(1)} x_k
+
b_j^{(1)}
\right).
\end{equation}

The second hidden layer computes

\begin{equation}
h_j^{(2)}
=
\mathrm{ReLU}
\left(
\sum_{k=1}^{32} w_{jk}^{(2)} h_k^{(1)}
+
b_j^{(2)}
\right).
\end{equation}

The output neuron predicts the next value of the coherence observable

\begin{equation}
\hat X(t_i)
=
\tanh
\left(
\sum_{j=1}^{16} w_j^{(3)} h_j^{(2)}
+
b^{(3)}
\right).
\end{equation}

The hyperbolic tangent activation function ensures that the prediction
remains within the physical range \([-1,1]\).

\begin{figure*}[t]
\centering
\includegraphics[width=0.95\linewidth]{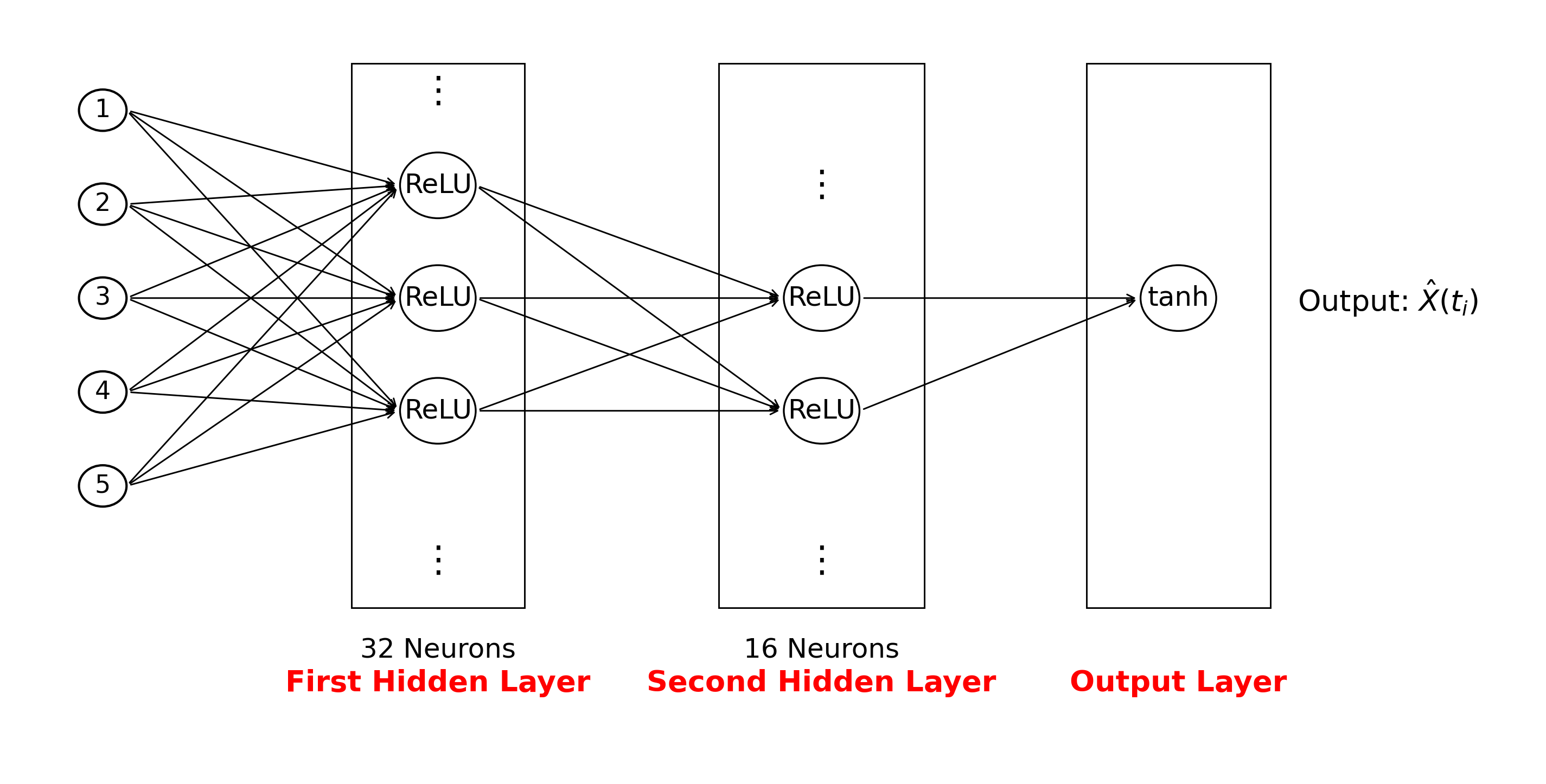}
\caption{Architecture of the neural network used for predicting the
future coherence observable \(X(t)\).
The input layer receives the five most recent measurements.
Two hidden layers perform nonlinear feature extraction,
and the output neuron produces the prediction \(\hat X(t_i)\).}
\label{fig:NN_architecture}
\end{figure*}

\subsection{Training Procedure}

The neural network parameters are optimized by minimizing the
mean–squared error between the predicted value and the true observable

\begin{equation}
\mathrm{MSE}
=
\frac{1}{N}
\sum_{i=1}^{N}
\left(
X(t_i)-\hat X(t_i)
\right)^2 .
\end{equation}

Optimization is performed using the Adam optimizer
\cite{Kingma2014adam}.

To evaluate generalization performance, the dataset is divided
sequentially so that the first \(50\%\) of the data are used for
training while the remaining \(50\%\) are reserved for testing.

\subsection{Numerical Simulation Framework}

The dataset used for training the machine learning model is generated
from numerical simulations of the open quantum system dynamics.

All simulations are performed using the Quantum Toolbox in Python
(QuTiP) \cite{qutip1,qutip2}.

The qubit evolves under the Hamiltonian

\begin{equation}
H(t)
=
\frac{\omega_0}{2}\sigma_z
+
\xi(t)\sigma_z ,
\end{equation}

where \(\xi(t)\) represents random telegraph noise.

The noise correlation function is

\begin{equation}
\langle\xi(t)\xi(s)\rangle
=
v^2 e^{-\kappa|t-s|}.
\end{equation}
\subsection{Simulation Parameters}

The simulations were performed with the following parameters

\begin{itemize}

\item total evolution time \(T=40\)

\item time step \(\Delta t=0.02\)

\item number of trajectories \(N_{\text{traj}}=3000\)

\item initial state
\[
|\psi(0)\rangle
=
\frac{|0\rangle+|1\rangle}{\sqrt{2}}
\]

\item sliding window size \(w=5\)

\end{itemize}

Two parameter regimes were considered:

Markovian regime

\[
v=1.0, \qquad \kappa=5.0
\]

Non-Markovian regime

\[
v=3.0, \qquad \kappa=0.25
\]
%%%%%%%%%%%%%%%%%%%%%%%%%%%%%%%%%%%%%%%%%
\section{Baseline RTN Dynamics and Prediction Validation}

In this section we establish the baseline coherence
dynamics of the qubit before introducing dynamical
decoupling. The purpose of this analysis is to validate
the physical RTN model together with the analytical and
machine-learning (ML) predictors in the absence of
control pulses.

Throughout this work, the system dynamics are monitored
through the transverse observable
$
X(t)=\langle\sigma_x(t)\rangle,
$
which quantifies the coherence of the qubit in the
$x$ basis. The qubit is initialized in the coherent
superposition state
$
|+\rangle=(|0\rangle+|1\rangle)/\sqrt{2},
$
corresponding to the initial condition
$
X(0)=1.
$
As the qubit interacts with the RTN environment,
stochastic phase accumulation progressively suppresses
the off-diagonal coherence terms, causing the observable
to decay toward zero.

In all simulations, the exact RTN dynamics are compared
with two predictive approaches:
(i) the analytical short-time predictor derived from the
Taylor expansion of the RTN dynamical equation, and
(ii) the ML predictor trained directly on short segments
of the coherence time series. This comparison provides a
direct validation of the predictive framework before the
introduction of adaptive dynamical decoupling.

\subsection{Stationary RTN}

We first consider the stationary RTN model, where the
switching rate $\kappa$ and coupling strength $v$ remain
constant throughout the evolution. In this case the
noise statistics are time independent, and the coherence
dynamics are governed by the competition between
environmental switching and coherent phase accumulation.

\begin{figure}[t]
\centering
\includegraphics[width=1.0\linewidth]{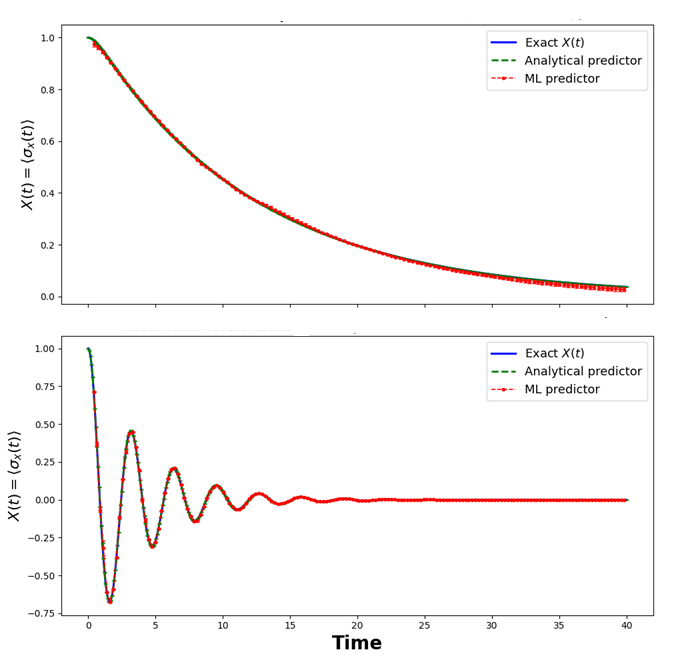}
\caption{
Coherence dynamics of a qubit interacting with a
stationary RTN environment.
The observable
$X(t)=\langle\sigma_x(t)\rangle$
is compared with the analytical predictor and the ML predictor.
The upper panel corresponds to the Markovian regime
($\kappa>v$), characterized by monotonic coherence
decay, while the lower panel corresponds to the
non-Markovian regime ($v\gtrsim\kappa$), where damped
oscillations and coherence revivals emerge.
Both predictors accurately reproduce the exact RTN
dynamics in the two regimes.
}
\label{fig:rtn_stationary}
\end{figure}

Figure~\ref{fig:rtn_stationary} shows the coherence
dynamics for representative Markovian and
non-Markovian stationary RTN regimes. In the Markovian
case ($\kappa>v$), rapid environmental switching
suppresses memory effects and produces a smooth
monotonic decay of coherence. In contrast, when the
coupling strength becomes comparable to or larger than
the switching rate ($v>\kappa$), the environment
retains memory over longer timescales, leading to
damped oscillations and coherence revivals that are
characteristic of non-Markovian dynamics.

In both regimes, the analytical predictor accurately
tracks the exact RTN solution, confirming the validity
of the local Taylor expansion for the coherence
evolution. The ML predictor also reproduces the exact
dynamics with high accuracy, demonstrating its ability
to learn both monotonic and oscillatory coherence
behavior directly from the time-series data.

\subsection{Non-stationary RTN}

We now extend the analysis to non-stationary
environments in which the RTN parameters become
time dependent,
$
\kappa\rightarrow\kappa(t)
$
and
$
v\rightarrow v(t).
$
The resulting coherence dynamics are governed by the
generalized non-stationary RTN equation derived earlier.
In this case the environmental properties evolve during
the system dynamics, leading to richer coherence
behavior than in the stationary regime.

Two representative models of non-stationary RTN are
considered. Model~1 describes localized Gaussian drift
of the environmental parameters, while Model~2
introduces persistent oscillatory modulation through
harmonic time dependence of both $\kappa(t)$ and
$v(t)$.

\begin{figure*}[t]
\centering
\includegraphics[width=0.9\textwidth]{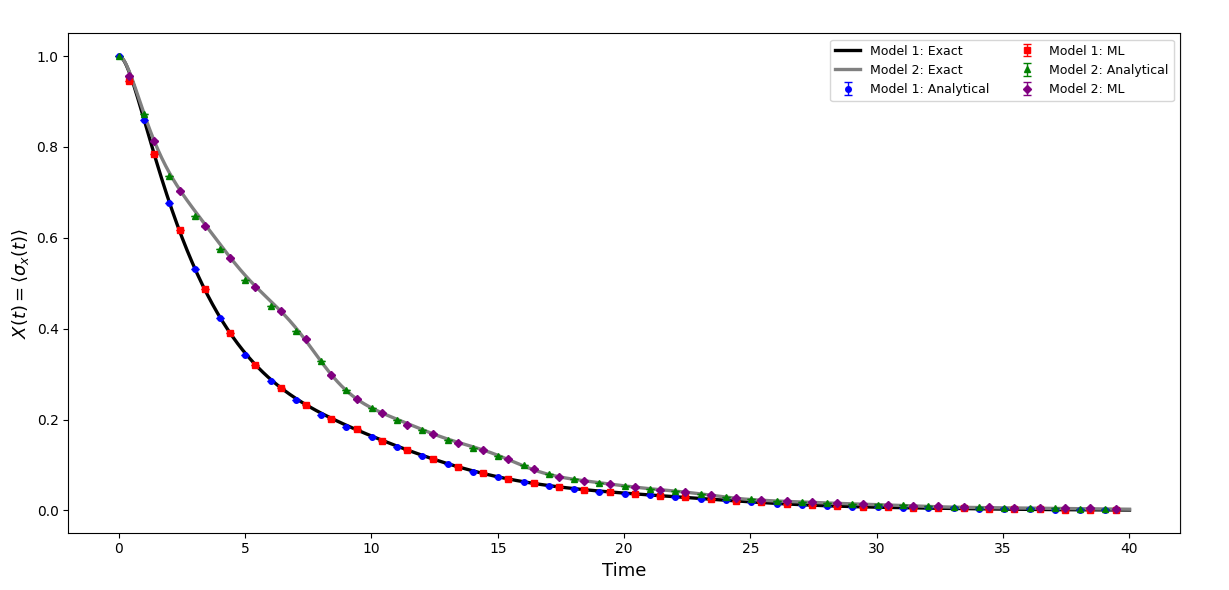}
\caption{
Coherence dynamics of a qubit interacting with a
non-stationary RTN environment in the Markovian regime.
The figure compares the exact coherence dynamics with
the analytical predictor and the ML predictor for
Model~1 (Gaussian drift) and Model~2 (oscillatory
modulation). Both predictors closely reproduce the exact
dynamics for the two non-stationary noise models.
}
\label{fig:nonstationary_markovian}
\end{figure*}
\begin{figure*}[t]
\centering
\includegraphics[width=0.9\textwidth]{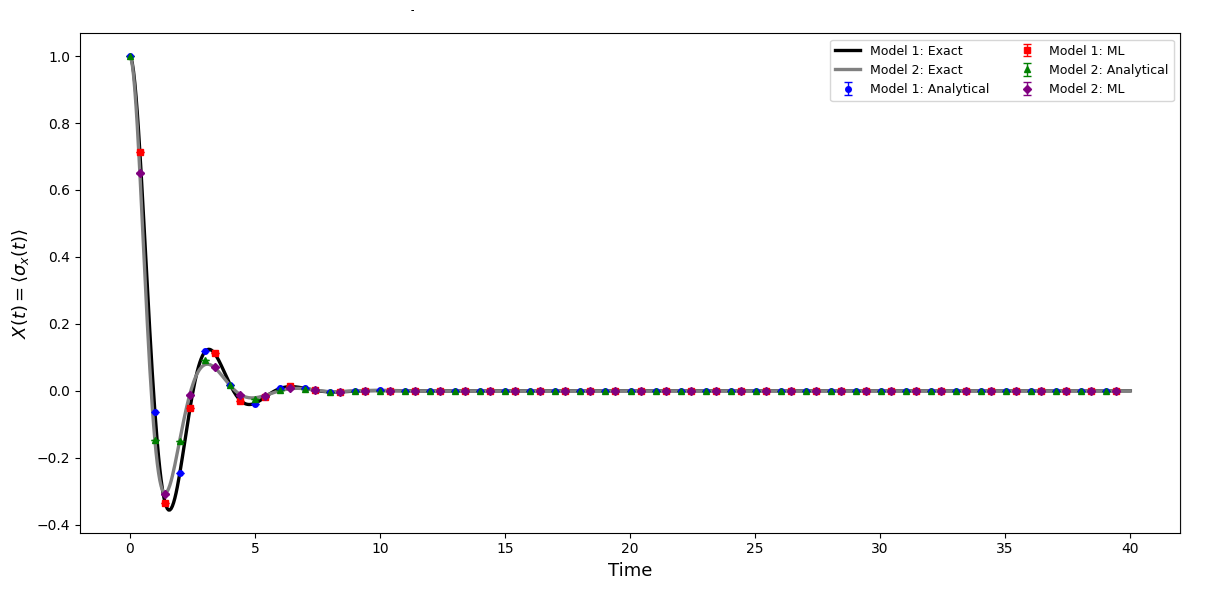}
\caption{
Coherence dynamics of a qubit interacting with a
non-stationary RTN environment in the non-Markovian
regime. The figure compares the exact coherence
dynamics with the analytical predictor and the ML
predictor for Model~1 (Gaussian drift) and Model~2
(oscillatory modulation). The damped oscillations and
coherence revivals reflect the presence of
non-Markovian memory effects.
}
\label{fig:nonstationary_nonmarkovian}
\end{figure*}

Figures~\ref{fig:nonstationary_markovian} and
\ref{fig:nonstationary_nonmarkovian} present the
coherence dynamics for the two non-stationary RTN
models in the Markovian and non-Markovian regimes,
respectively. In all cases, the analytical predictor
derived from the non-stationary RTN equation remains in
excellent agreement with the exact dynamics despite the
time dependence of the environmental parameters. The ML
predictor likewise reconstructs the coherence evolution
with high accuracy directly from the observable time
series.

For Model~1, the localized Gaussian drift produces
smooth temporal variations in the coherence decay rate,
whereas Model~2 generates more structured dynamics due
to the persistent oscillatory modulation of the RTN
parameters. In the non-Markovian regime, these
time-dependent variations enhance the appearance of
coherence oscillations and revivals associated with
environmental memory effects.

Overall, the close agreement between the exact RTN
dynamics, the analytical predictor, and the ML model
demonstrates that both predictive approaches reliably
forecast the short-time evolution of the system even in
the presence of strongly time-dependent noise.
Throughout the remainder of this work, Model~1
(Gaussian drift) will be adopted as the reference model
for the non-stationary environment.
%%%%%%%%%%%%%%%%%%%%%%%%%%%%%%%%%%%%%%%%%%

\section{Dynamical Decoupling Control}
\label{sec:DD}

Dynamical decoupling (DD) is a widely used technique for protecting
quantum coherence from environmental noise by applying sequences of
control pulses that effectively average out the system--environment
interaction \cite{Viola1998,Viola1999,Khodjasteh2005,Uhrig2007,Suter2016}.
The central idea is to apply rapid control operations to the qubit so
that the accumulated phase induced by environmental fluctuations
cancels over time.

In the open quantum system model introduced in Sec.~\ref{sec2},
the qubit interacts with the environment through the interaction
Hamiltonian

\begin{equation}
H_{SE}(t)=\xi(t)\sigma_z ,
\end{equation}

where $\xi(t)$ denotes the stochastic noise field generated by the
environment. Because the interaction operator is proportional to
$\sigma_z$, the noise does not induce transitions between the energy
levels of the qubit. Instead, it produces fluctuations in the relative
phase between the basis states, leading to pure dephasing.

During free evolution, the stochastic fluctuations of $\xi(t)$ produce
random phase shifts that gradually destroy the coherence of the qubit.
This loss of coherence can be monitored through the transverse
observable

\begin{equation}
X(t)=\langle\sigma_x(t)\rangle ,
\end{equation}

which is directly related to the off--diagonal element of the density
matrix as discussed in the previous section.

Dynamical decoupling suppresses this decoherence by applying control
pulses that reverse the sign of the system--environment interaction.
In particular, a $\pi$ pulse applied around the $x$ axis transforms the
operator $\sigma_z$ according to

\begin{equation}
\sigma_x \sigma_z \sigma_x = -\sigma_z .
\end{equation}

Therefore, every time a $\pi$ pulse is applied, the sign of the
interaction Hamiltonian is inverted. As a result, the phase accumulated
after the pulse cancels the phase accumulated before the pulse.

If such pulses are applied repeatedly during the system evolution, the
alternating sign of the interaction averages the environmental
fluctuations and suppresses the accumulation of stochastic phase.
When the pulse interval is shorter than the correlation time of the
environment, this averaging effect can significantly reduce decoherence.

The action of the pulse sequence can be described using a switching
function $y(t)$ defined as

\begin{equation}
y(t)=(-1)^{n(t)},
\end{equation}

where $n(t)$ is the number of $\pi$ pulses applied up to time $t$.
The effective interaction Hamiltonian under dynamical decoupling
therefore becomes

\begin{equation}
H_{SE}^{(\mathrm{DD})}(t)=y(t)\,\xi(t)\sigma_z .
\end{equation}

Thus, every control pulse flips the sign of the interaction Hamiltonian,
which causes the accumulated phase noise to alternate in sign and
reduces the net dephasing experienced by the qubit.

\begin{figure}[t]
\centering
\includegraphics[width=1.05\linewidth]{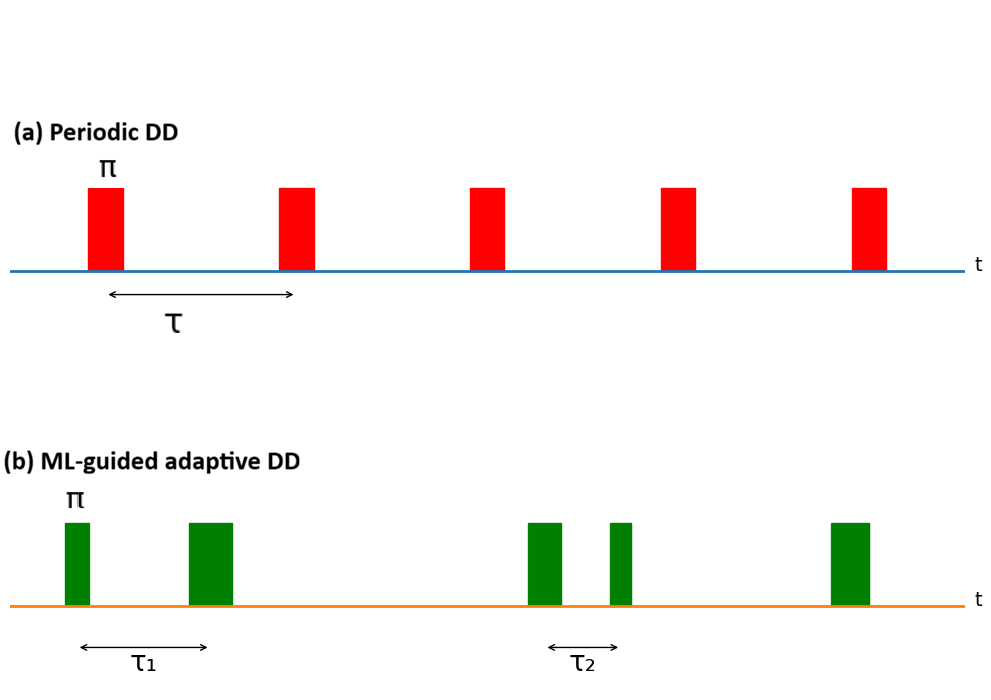}
\caption{Conceptual illustration of dynamical decoupling.
(a) During free evolution environmental fluctuations produce random
phase shifts that gradually destroy the coherence of the qubit.
(b) When $\pi$ pulses are applied periodically, the sign of the
interaction Hamiltonian is repeatedly inverted, causing the accumulated
phase noise to partially cancel and leading to improved coherence
preservation.}
\label{fig:dd_schematic}
\end{figure}

\subsection{Periodic Dynamical Decoupling}

The simplest implementation of dynamical decoupling is periodic
dynamical decoupling (PDD), in which $\pi$ pulses are applied at regular
time intervals throughout the evolution of the system
\cite{Viola1998,Viola1999,Khodjasteh2005,Uhrig2007,Cywinski2008,Suter2016}.

Let $\tau$ denote the time interval between consecutive pulses and
$T$ the total evolution time. The pulses are then applied at the times

\begin{equation}
t_j=j\tau , \qquad j=1,2,\dots,N,
\end{equation}

where the total number of pulses is

\begin{equation}
N=\frac{T}{\tau}.
\end{equation}

Between two pulses the system evolves freely under the noisy Hamiltonian.
Each pulse flips the qubit state and reverses the sign of the
system--environment interaction. As a result, the phase accumulated from
environmental fluctuations alternates in sign during successive time
intervals.

When the pulse spacing $\tau$ is sufficiently small compared with the
correlation time of the environment, the alternating phase contributions
partially cancel each other, leading to a suppression of decoherence.
For a fixed total evolution time $T$, decreasing the pulse spacing
(i.e., increasing the number of pulses $N$) generally improves the
averaging of environmental fluctuations and enhances coherence
preservation.

However, periodic dynamical decoupling assumes that the statistical
properties of the environment remain stationary during the evolution.
If the noise characteristics vary in time, the optimal pulse locations
may change dynamically and fixed pulse schedules may become suboptimal.

\subsection{Machine-Learning Adaptive Dynamical Decoupling}

To overcome the limitations of fixed pulse schedules, we introduce an
adaptive dynamical decoupling protocol guided by machine learning.
Instead of applying pulses at predetermined intervals, the pulse
locations are determined using predictions of the future coherence
dynamics.

The machine-learning model predicts the short-time evolution of the
coherence observable using a sliding window of previously measured
values. Let $w$ denote the size of the input window. The predicted
future coherence is

\begin{equation}
\hat X(t_{i+1})
=
M\big(X(t_{i-w+1}),\ldots,X(t_i)\big),
\end{equation}

where $M(\cdot)$ represents the trained machine-learning predictor.

The predicted coherence trajectory is then used to determine whether a
control pulse should be applied. In particular, a pulse is triggered
whenever the predicted decrease in coherence exceeds a threshold
$\epsilon$,

\begin{equation}
u_i=
\begin{cases}
1 & \hat X(t_{i+1})-\hat X(t_i)<-\epsilon,\\
0 & \text{otherwise}.
\end{cases}
\end{equation}

Here $u_i=1$ indicates that a $\pi$ pulse is applied at time $t_i$,
while $u_i=0$ corresponds to free evolution without control.

The resulting sequence of pulses defines a switching function

\begin{equation}
y_{\mathrm{ML}}(t)=(-1)^{\sum_i u_i},
\end{equation}

which determines the sign of the effective interaction Hamiltonian
under control. Each time a pulse is applied, the sign of the interaction
is inverted, reversing the accumulated phase noise.

In contrast to periodic dynamical decoupling, the total number of pulses
in the adaptive protocol is not fixed in advance. Instead, pulses are
applied only when the predicted coherence decay exceeds the threshold
$\epsilon$. As a result, the pulse sequence adapts dynamically to the
instantaneous behavior of the environment.

This predictive control strategy is particularly advantageous in
non-stationary environments where the parameters of the noise process
change in time. In such situations, periodic pulse schedules may fail
to suppress decoherence efficiently, whereas the machine-learning-guided
protocol can adapt the pulse locations according to the predicted
evolution of the coherence observable.
%%%%%%%%%%%%%%%%%%%%%%%%%%%%%%%%%

%%%%%%%%%%%%%%%%%%%%%%%%%%%%%%%%%%%%%%%%%%
\section{Results}\label{sec:results}

In this section we present the numerical results for the coherence dynamics of a single qubit interacting with a random telegraph noise (RTN) environment and investigate the performance of dynamical decoupling protocols for suppressing decoherence. 
The goal is to compare the effectiveness of conventional periodic dynamical decoupling sequences with the proposed machine-learning-guided adaptive dynamical decoupling protocol.

Throughout this work the coherence of the qubit is quantified using the expectation value $
X(t)=\langle\sigma_x(t)\rangle $,

which directly measures the preservation of quantum coherence in the $x$ basis. 
For an initial state aligned along the $x$ direction this observable is proportional to the real part of the off-diagonal density matrix element $\rho_{01}(t)$ and therefore provides a direct indicator of decoherence in the system.

We consider both stationary and non-stationary RTN environments. 
In the stationary case the statistical properties of the noise remain constant in time, while in the non-stationary case the environmental parameters vary dynamically during the evolution. 
Because the behavior of dynamical decoupling protocols depends strongly on the temporal structure of the noise, it is important to analyze these two cases separately.
\subsection{Stationary Random Telegraph Noise}

We first analyze the stationary random telegraph noise (RTN) model, in which the environmental parameters remain constant during the evolution. In this scenario the environment is characterized by a fixed switching rate $\kappa$ and a constant noise amplitude $v$. The fluctuating environment randomly switches between the two values $\pm v$, generating exponentially correlated noise with a Lorentzian power spectrum.

The dynamical behavior of the system depends on the relative magnitude of the switching rate and the noise amplitude. When the switching rate dominates ($\kappa > v$), the environment fluctuates rapidly and the system effectively experiences memoryless noise, leading to Markovian dynamics. Conversely, when the noise amplitude exceeds the switching rate ($v > \kappa$), the environment switches slowly and retains memory of its previous state, producing non-Markovian dynamics characterized by oscillations and partial revivals of coherence.

\begin{figure*}[t]
\centering
\includegraphics[width=1.0\linewidth]{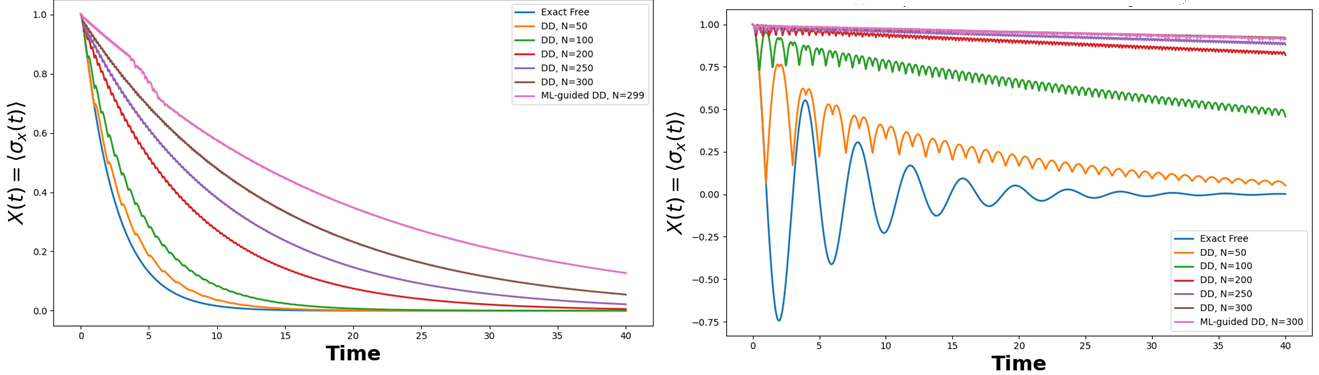}
\caption{Coherence dynamics $X(t)=\langle\sigma_x(t)\rangle$ for a qubit subjected to stationary random telegraph noise. 
Left panel: Markovian regime ($\kappa>v$), where the environment fluctuates rapidly and the coherence decays monotonically. 
Right panel: Non-Markovian regime ($v>\kappa$), where environmental memory produces oscillations and partial revivals of coherence. 
The blue curve corresponds to free evolution without control. 
Colored curves correspond to periodic dynamical decoupling sequences with different numbers of pulses, while the dashed curve with markers represents the ML-guided adaptive dynamical decoupling protocol.}
\label{fig:stationary_rtn_results}
\end{figure*}

Figure~\ref{fig:stationary_rtn_results} illustrates the coherence dynamics
$X(t)=\langle\sigma_x(t)\rangle$ for the stationary RTN model in both regimes. The left panel corresponds to the Markovian regime, while the right panel shows the non-Markovian regime. In both cases the blue curve represents the free evolution of the qubit in the absence of control pulses. The remaining curves correspond to periodic dynamical decoupling (DD) sequences with different numbers of $\pi$ pulses, while the ML-guided adaptive DD protocol is shown for comparison.

In the stationary Markovian regime (left panel), the free evolution exhibits a smooth monotonic decay of coherence due to rapid environmental fluctuations. Applying periodic dynamical decoupling significantly suppresses this decoherence by repeatedly reversing the sign of the system–environment interaction. As the number of pulses increases, the interaction is more effectively averaged out and the coherence decay becomes progressively slower.

The ML-guided adaptive DD protocol can provide a modest improvement in this regime. Although the noise statistics remain stationary, the adaptive algorithm identifies time intervals where the predicted coherence loss is strongest and allocates pulses accordingly. This nonuniform distribution of pulses can yield slightly improved coherence preservation compared with strictly periodic pulse sequences when the number of available pulses is limited.

The behavior is qualitatively different in the stationary non-Markovian regime (right panel). In this case the environment switches slowly, so the system retains memory of its previous interactions with the environment. As a result, the free evolution exhibits pronounced oscillations and partial revivals of coherence, which correspond to information temporarily flowing back from the environment to the system.

Periodic dynamical decoupling is particularly effective in this regime. Because the environmental fluctuations are slow and correlated in time, uniformly spaced $\pi$ pulses can efficiently refocus the phase accumulated during each noise interval. Increasing the number of pulses strongly suppresses both the oscillations and the overall decay of coherence, as clearly observed in Fig.~\ref{fig:stationary_rtn_results}.

Since the stationary environment has fixed statistical properties, the optimal pulse spacing remains approximately uniform throughout the evolution. Consequently, periodic dynamical decoupling already provides near-optimal suppression of decoherence, and the advantage of adaptive pulse placement becomes relatively modest in this stationary scenario.

The number of dynamical decoupling pulses used in our simulations is chosen to be consistent with experimentally achievable pulse sequences reported in current quantum platforms. Representative pulse durations and typical DD sequence lengths are summarized in Table~\ref{tab:dd_experimental}.

\begin{table*}[t]
\centering
\begin{tabular}{lccc}
\hline
Quantum Platform & $\pi$-Pulse Duration & Typical DD Pulses & References \\
\hline

Superconducting qubits 
& $10$--$50$ ns 
& $10$--$200$ 
& \cite{Bylander2011,Yan2013} \\

NV centers in diamond 
& $20$--$100$ ns 
& $100$--$1000$ 
& \cite{Suter2016,Biercuk2009} \\

Trapped-ion qubits 
& $1$--$10$ $\mu$s 
& $10$--$100$ 
& \cite{Biercuk2009} \\

Semiconductor quantum dots 
& $10$--$100$ ns 
& $50$--$500$ 
& \cite{Medford2012,Bluhm2011} \\

\hline
\end{tabular}

\caption{
Typical $\pi$-pulse durations and experimentally achievable dynamical decoupling sequence lengths for several quantum computing platforms. 
The pulse numbers used in this work ($N=20$--$160$) lie within the experimentally accessible range reported in current quantum hardware.
}
\label{tab:dd_experimental}
\end{table*}
\subsection{Non-stationary Random Telegraph Noise}

We now extend the analysis to the more realistic case of
\emph{non-stationary} random telegraph noise (RTN), in which the
environmental parameters vary slowly during the evolution.
In realistic solid-state quantum devices, the microscopic fluctuators
responsible for decoherence are not perfectly stable.
Their coupling strengths and switching rates may drift due to thermal
fluctuations, charge trapping, slow defects in the substrate, or
changes in the local electromagnetic environment.
As a result, the qubit does not experience a noise process with fixed
statistical properties, but rather a time-dependent decoherence
mechanism whose effective timescale changes throughout the evolution.

In our simulations, the non-stationary RTN parameters are allowed to
vary within the ranges $
0.5 \le v(t) \le 1.5$ and $
0.1 \le \kappa(t) \le 1.0$.
These values are consistent with experimentally relevant fluctuator
strengths and switching rates reported for solid-state quantum
platforms, including superconducting qubits and semiconductor spin
qubits, where RTN and related bistable defect noise have been widely
observed~\cite{Paladino2014,Galperin2006,Bergli2009}.

Because both $v(t)$ and $\kappa(t)$ vary in time, the dynamical regime
of the open system may also change during the evolution.
When the instantaneous switching rate is sufficiently large compared
with the noise amplitude, the environment behaves effectively
Markovian and the coherence decays in a smoother and more monotonic
way.
By contrast, when the coupling becomes comparable to or stronger than
the switching process, environmental memory becomes more important and
the coherence exhibits oscillatory non-Markovian features.
This time dependence makes pulse design substantially more difficult
than in the stationary case, since the optimal control interval is no
longer constant.

\begin{figure*}[t]
\centering
\includegraphics[width=\linewidth]{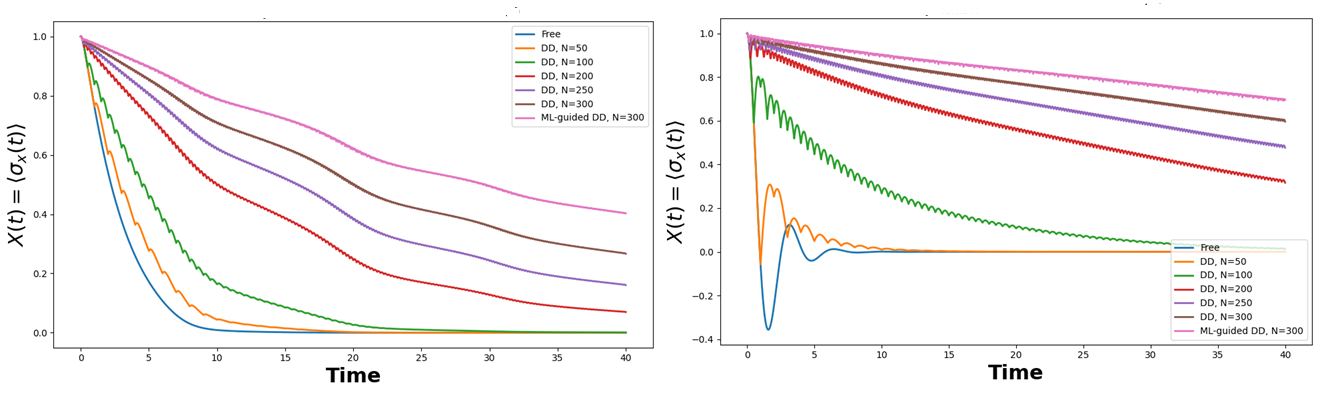}
\caption{
Coherence dynamics $X(t)=\langle\sigma_x(t)\rangle$ for a qubit subjected to
non-stationary random telegraph noise.
The RTN parameters vary in time within the ranges
$0.5 \le v(t) \le 1.5$ and $0.1 \le \kappa(t) \le 1.0$.
(\textbf{Left}) Non-stationary Markovian regime.
The free evolution decays rapidly, while periodic DD improves coherence
as the number of pulses increases.
The ML-guided adaptive DD protocol preserves the coherence more
efficiently by applying pulses at times where decoherence is predicted
to be strongest.
(\textbf{Right}) Non-stationary non-Markovian regime.
The free evolution shows oscillatory behavior associated with
environmental memory.
Periodic DD suppresses the oscillations and slows the decay, but the
adaptive ML-guided protocol maintains the highest coherence throughout
the evolution.
In both panels the same pulse numbers are used as in the stationary
case, $N=50,100,200,250,300$, allowing a direct comparison between
fixed periodic control and adaptive pulse placement.
}
\label{fig:nonstationary_rtn}
\end{figure*}

Figure~\ref{fig:nonstationary_rtn} presents the coherence dynamics $
X(t)=\langle \sigma_x(t)\rangle $
for a qubit subjected to non-stationary RTN.
The left panel corresponds to the non-stationary Markovian regime,
while the right panel corresponds to the non-stationary
non-Markovian regime.
In order to make the comparison with the stationary case transparent,
we use the same pulse numbers as in the stationary analysis,
namely $N=50,100,200,250,300$.
In each panel we compare three scenarios:
(i) free evolution without control,
(ii) periodic dynamical decoupling (DD) with fixed pulse spacing, and
(iii) machine-learning-guided adaptive DD.

In the non-stationary Markovian regime shown in the left panel of
Fig.~\ref{fig:nonstationary_rtn}, the free evolution still decays
rapidly, but unlike the stationary case, the decay rate is not governed
by a single fixed timescale.
Instead, the effective decoherence strength changes as
$v(t)$ and $\kappa(t)$ drift.
Periodic DD therefore remains helpful, but its effectiveness is limited
by the fact that its pulse spacing is fixed in advance.
A pulse interval that is suitable at one stage of the evolution may no
longer be optimal later when the instantaneous noise timescale changes.
This explains why even dense periodic sequences eventually become less
efficient than a control strategy that can adapt to the evolving
environment.

The ML-guided adaptive DD protocol overcomes this limitation by using
the short-time behavior of the observable $X(t)$ to infer when the next
significant coherence loss is likely to occur.
Rather than applying pulses uniformly in time, it determines
\emph{when a pulse is most needed}.
As a result, the ML-guided protocol preserves higher coherence than the
regular DD sequences even when the same pulse budget is considered.
This is the key advantage of adaptive control in a non-stationary
Markovian environment: the model identifies the changing effective
decoherence timescale and adjusts the pulse placement accordingly.

The advantage of adaptive DD becomes even more significant in the
non-stationary non-Markovian regime shown in the right panel of
Fig.~\ref{fig:nonstationary_rtn}.
Here the coherence dynamics exhibit oscillatory structure due to the
combined effects of time-dependent parameters and environmental memory.
In this regime, the system is influenced not only by the instantaneous
noise strength but also by the history of the qubit--environment
interaction.
Consequently, a fixed periodic pulse sequence cannot fully exploit the
structure of the dynamics.
Although periodic DD still improves the coherence and increasingly
suppresses the oscillations as $N$ grows, it remains fundamentally
limited by its inability to respond to the changing locations and
strengths of the memory-induced coherence losses.

By contrast, the ML-guided protocol learns the short-time evolution of
$X(t)$ and can anticipate the approach of these loss regions.
This predictive capability allows the adaptive controller to place
pulses at the most relevant times, rather than distributing them
uniformly.
The result is a substantially better preservation of coherence across
the entire evolution.
Importantly, this enhancement is achieved while using the same pulse
numbers as those employed in the stationary analysis, which makes the
comparison fair and highlights that the improvement comes from
\emph{pulse timing}, not simply from applying more pulses.

These results clearly demonstrate the practical importance of adaptive
dynamical decoupling in realistic time-dependent environments.
In stationary noise, periodic DD can already approach the optimal
behavior because the noise timescale is fixed.
In non-stationary RTN, however, the environment evolves during the
experiment, and fixed pulse schedules become intrinsically suboptimal.
The ML-guided strategy provides a direct answer to this limitation:
it reveals \emph{when} the pulses should be applied.
This ability to infer pulse timing from the observed qubit dynamics is
precisely what makes adaptive DD especially valuable for realistic
quantum hardware, where noise properties are not constant in time.
\subsection{Quantifying Coherence Preservation}

A central objective of this work is to determine how effectively
different control strategies preserve quantum coherence in the
presence of environmental noise.
While the coherence dynamics are illustrated in the previous figures,
a meaningful comparison between different control protocols requires
a single quantitative measure that captures the overall coherence
retained during the entire evolution.

Instead of evaluating the coherence at a single instant of time, it is
more informative to quantify how much coherence is preserved throughout
the full duration of the experiment.
This approach reflects the operational goal of quantum control:
a useful quantum device must maintain coherence continuously during its
operation rather than only at isolated moments.

To quantify this behavior we consider the transverse qubit coherence

\begin{equation}
X(t)=\langle\sigma_x(t)\rangle,
\end{equation}

which measures the off-diagonal elements of the qubit density matrix
and therefore directly characterizes the degree of quantum coherence in
the system.

Since different control protocols may operate on different scales,
we introduce the normalized coherence magnitude

\begin{equation}
C(t)=\frac{|X(t)|}{|X(0)|}.
\end{equation}

This normalization ensures that $C(0)=1$ for all protocols and allows a
direct comparison between different control strategies.
The absolute value is used because in non-Markovian regimes the
coherence can oscillate and change sign; taking the magnitude avoids
artificial cancellation when integrating the signal over time.

To evaluate the total coherence preserved during the experiment we
introduce the time-integrated coherence

\begin{equation}
\mathcal{A}[C]=\int_0^T C(t)\,dt .
\end{equation}

Physically, $\mathcal{A}[C]$ corresponds to the area under the
coherence curve and therefore measures the cumulative coherence
retained by the system over the entire time interval $[0,T]$.
A larger value of $\mathcal{A}[C]$ indicates that the qubit maintains
coherence for a longer portion of the evolution and thus reflects a
more effective control protocol.
Time-integrated coherence measures are widely used in the analysis of
dynamical decoupling and noise-suppression strategies in open quantum
systems~\cite{Cywinski2008,Paladino2014}.

The values of $\mathcal{A}[C]$ obtained in our simulations are
summarized in Table~\ref{tab:integrated_coherence_all}.
The table compares three scenarios: free evolution without control,
the best periodic dynamical decoupling (DD) sequence, and the
machine-learning-guided adaptive DD protocol.
The periodic DD results correspond to the best performance obtained
among the tested pulse numbers
$N=50,100,200,250,300$.

\begin{table*}[t]
\centering
\caption{Integrated coherence 
$\mathcal{A}[C]=\int_0^T C(t)\,dt$
for different noise regimes. 
The best periodic DD result is selected from the tested pulse numbers
$N=50,100,200,250,300$.}
\begin{tabular}{lcccc}
\hline
Protocol &
\makecell{Stationary \\ Markovian} &
\makecell{Stationary \\ Non-Markovian} &
\makecell{Non-stationary \\ Markovian} &
\makecell{Non-stationary \\ Non-Markovian} \\
\hline

Free evolution 
& 2.50 & 4.28 & 2.84 & 1.085 \\

Best periodic DD 
& 12.90 & 38.16 & 21.93 & 31.07 \\

ML-guided DD 
& 15.86 & 38.15 & 26.00 & 33.38 \\

ML advantage (\%)
& 22.9\% & $\approx 0\%$ & 18.6\% & 7.4\% \\

\hline
\end{tabular}
\label{tab:integrated_coherence_all}
\end{table*}

Table~\ref{tab:integrated_coherence_all} provides a global comparison of
coherence preservation across the four noise regimes considered in this
work.
The first row corresponds to free evolution, which represents the
baseline case where no control pulses are applied.
As expected, the integrated coherence values are relatively small,
indicating rapid loss of coherence due to environmental noise.

The second row reports the best performance achieved by periodic
dynamical decoupling.
In all regimes periodic DD dramatically improves coherence preservation
compared with free evolution.
For example, in the stationary Markovian regime the integrated
coherence increases from $2.50$ to $12.90$, demonstrating the strong
ability of pulse sequences to suppress decoherence by averaging out the
system--environment interaction.

The third row shows the performance of the machine-learning-guided
adaptive DD protocol.
This protocol uses short-time information from the observed qubit
dynamics to predict when coherence loss is likely to occur and applies
control pulses at the most effective times rather than distributing
them uniformly.
The resulting integrated coherence values are higher than those
obtained with periodic DD in most regimes.

The last row of the table highlights the relative advantage of the
adaptive protocol over the best periodic DD sequence.
In the stationary Markovian regime the ML-guided protocol achieves a
$22.9\%$ improvement in the total preserved coherence.
This demonstrates that even when the environment is memoryless,
adaptive pulse placement can still outperform fixed periodic control.

In the stationary non-Markovian regime, however, the improvement is
negligible.
Both periodic DD and the ML-guided protocol yield nearly identical
integrated coherence values.
This occurs because the environmental properties remain constant in
time, allowing periodic pulse sequences to already approach the optimal
control strategy.

The largest advantage of the adaptive protocol appears in the
non-stationary Markovian regime.
Here the environmental parameters drift in time, causing the
decoherence rate to vary during the evolution.
In this situation periodic DD becomes intrinsically suboptimal because
its pulse spacing is fixed in advance.
By contrast, the ML-guided protocol dynamically adjusts the pulse
timing based on the observed coherence dynamics, resulting in an
additional $18.6\%$ improvement over periodic DD.

Finally, in the non-stationary non-Markovian regime the environment
exhibits both temporal drift and memory effects.
Periodic DD already provides substantial coherence protection, but the
ML-guided protocol still achieves a further $7.4\%$ enhancement by
anticipating the locations of coherence loss associated with
memory-induced fluctuations.

Overall, Table~\ref{tab:integrated_coherence_all} clearly demonstrates
that adaptive control provides its greatest advantage when the noise
environment is non-stationary.
In realistic quantum devices, where noise properties often drift in
time, such adaptive strategies therefore offer a promising route toward
more robust preservation of quantum coherence.

A direct comparison between the stationary results in Fig.~\ref{fig:stationary_rtn_results} and the
non-stationary results in Fig.~\ref{fig:nonstationary_rtn}  clearly reveals where the
machine-learning-assisted adaptive dynamical decoupling protocol
provides its greatest advantage.
In the stationary regimes, particularly for stationary non-Markovian
noise, periodic DD already performs relatively efficiently because the
environmental timescales remain fixed throughout the evolution.
As a result, a properly chosen periodic pulse spacing can remain close
to optimal over long times.

In contrast, the non-stationary non-Markovian regime shown in
Fig.~\ref{fig:nonstationary_rtn} represents a far more challenging physical situation.
Here the environment simultaneously exhibits temporal drift,
time-dependent decoherence rates, coherence revivals, and memory
effects.
Consequently, the optimal pulse locations evolve continuously during
the dynamics and cannot be captured efficiently by fixed periodic pulse
sequences.
This is precisely where the predictive machine-learning-assisted
framework becomes most effective.

Unlike periodic DD, the adaptive DD+ML protocol continuously learns
the short-time evolution of the coherence observable directly from the
observed qubit dynamics and predicts when the next significant
coherence loss is likely to occur.
This allows the controller to dynamically reposition pulses according
to the instantaneous structure of the environment rather than applying
them uniformly in time.
The improvement becomes especially visible in Fig.~10, where the
adaptive protocol maintains substantially higher coherence throughout
the evolution despite using the same pulse budget as the periodic DD
sequences.

Therefore, the comparison between Figs.~\ref{fig:stationary_rtn_results} and~\ref{fig:nonstationary_rtn} demonstrates a
central result of this work: the advantage of predictive
machine-learning-assisted dynamical decoupling grows significantly as
the environment becomes both non-Markovian and non-stationary.
These are precisely the realistic noise conditions under which
conventional fixed pulse protocols become intrinsically limited, while
adaptive predictive control remains highly effective.
\section{Conclusion}\label{sec:conclusion}

In this work we developed a predictive machine-learning-assisted adaptive dynamical decoupling framework for protecting quantum coherence in realistic noisy quantum environments. By combining analytical open-system modeling with data-driven forecasting, the proposed framework dynamically predicts short-time coherence evolution and adaptively determines when control pulses should be applied. Unlike conventional periodic dynamical decoupling protocols based on fixed pulse schedules, the present approach continuously responds to the instantaneous behavior of the quantum system and the evolving characteristics of the environment.

Using random telegraph noise as a physically motivated model of environmental fluctuations, we investigated coherence preservation in stationary and non-stationary environments spanning both Markovian and non-Markovian regimes. Two distinct models of non-stationary noise were considered, including localized Gaussian environmental drift and oscillatory temporal modulation of the noise parameters. These models capture realistic scenarios in which environmental properties evolve continuously during the system dynamics, producing conditions under which conventional fixed pulse protocols become intrinsically inefficient.

Our results demonstrate that the predictive machine-learning-assisted protocol consistently outperforms conventional periodic dynamical decoupling while using a comparable number of control pulses. The enhancement becomes especially pronounced in non-Markovian and non-stationary environments, where memory effects, coherence revivals, information backflow, and temporally evolving noise strongly reduce the effectiveness of static pulse schedules. In such regimes, the adaptive framework continuously learns the changing coherence dynamics directly from the observed qubit evolution and anticipates future decoherence events before they occur. This predictive capability enables the application of control pulses at dynamically optimal times, leading to substantially improved coherence preservation compared with conventional approaches.

A central result of this work is that predictive control becomes increasingly advantageous precisely in the most difficult and experimentally relevant noise environments, namely those characterized by temporal drift, evolving spectral properties, and non-Markovian memory effects. While fixed dynamical decoupling sequences are designed under the assumption of stationary noise, realistic quantum devices rarely operate under perfectly stationary conditions. The present framework overcomes this limitation by introducing a fully adaptive strategy capable of responding in real time to complex environmental fluctuations.

Beyond the specific random telegraph noise model considered here, the framework developed in this work provides a general paradigm for intelligent quantum control in open quantum systems. The integration of machine learning with dynamical decoupling establishes a scalable route toward autonomous noise-adaptive quantum error suppression, with potential applications in superconducting circuits, spin qubits, trapped ions, quantum sensors, and near-term quantum processors operating in realistic noisy environments.

Overall, the present results demonstrate that predictive machine-learning-assisted dynamical decoupling is not merely an incremental improvement over conventional pulse protocols, but rather represents a fundamentally new direction for adaptive quantum control. By combining predictive intelligence with real-time coherence protection, this approach opens a pathway toward robust and scalable quantum technologies capable of operating reliably in complex dynamical environments.
\section*{Acknowledgement}
L.-A.W. is supported by the Basque Country Government (Grant No.\ IT1470-22) and Grant No.\ PGC2018-101355-B-I00 funded by MCIN/AEI/10.13039/501100011033, the Ministry for Digital Transformation and Civil Service of the Spanish Government through the QUANTUM ENIA project call—Quantum Spain project, and by the European Union through the Recovery, Transformation and Resilience Plan-NextGenerationEU within the framework of the Digital Spain 2026 Agenda.

A.~Abu-Nada acknowledges the support of Sharjah Maritime Academy for covering the publication fees associated with this work.

%%%%%%%%%%%%%%%%%%%%%%%%%%%%%%%%%%%%

\bibliography{main}

\end{document}